\documentclass[a4paper,10pt]{article}
\usepackage{arxiv}

\usepackage{amsmath}
\usepackage{amsfonts}
\usepackage{mathtools}
\usepackage{hyperref}       % hyperlinks
\usepackage{url}            % simple URL typesetting

\usepackage{array}

\usepackage{tikz} 
\usetikzlibrary{spy}

\usepackage[T1]{fontenc}
\usepackage{graphicx}
\newcommand{\x}{\mathbf{x}}
\newcommand{\y}{\mathbf{y}}
\newcommand{\z}{\mathbf{z}}

\newcommand{\off}{\mathbf{o}}
\newcommand{\pat}{\mathbf{p}}
\newcommand{\one}{\mathbf{1}}

\newcommand{\id}{\mathrm{Id}}
\newcommand{\D}{\mathcal{D}}
\renewcommand{\O}{\mathcal{O}}
\renewcommand{\P}{\mathcal{P}}
\newcommand{\W}{\mathcal{W}}
\newcommand{\DS}{\mathtt{DS}}

\newcommand{\N}{\mathcal{N}}

\newcommand{\U}{\mathcal{U}}
\renewcommand{\L}{\mathcal{L}}
\newcommand{\R}{\mathbb{R}}
\newcommand{\E}{\mathbb{E}}
\newcommand{\NA}{\mathrm{NA}}
\newcommand{\LD}{\mathrm{LD}}
\newcommand{\SD}{\mathrm{SD}}
\newcommand{\metad}[2]{\ensuremath{(#1 \ #2)^\top}}

\newcommand{\site}[1]{
    \setbox0=\hbox{#1}%
    \dimen0\wd0%
    \divide\dimen0 by 10%
    \begin{tikzpicture}[baseline=(a.base)]%
        \useasboundingbox (-\the\dimen0,0pt) rectangle (\the\dimen0,.5pt);
        \node[circle,draw,outer sep=0pt,inner sep=0.1ex] (a) {#1};
    \end{tikzpicture}
}

\makeatletter
\newcommand{\subalign}[1]{%
  \vcenter{%
    \Let@ \restore@math@cr \default@tag
    \baselineskip\fontdimen10 \scriptfont\tw@
    \advance\baselineskip\fontdimen12 \scriptfont\tw@
    \lineskip\thr@@\fontdimen8 \scriptfont\thr@@
    \lineskiplimit\lineskip
    \ialign{\hfil$\m@th\scriptstyle##$&$\m@th\scriptstyle{}##$\hfil\crcr
      #1\crcr
    }%
  }%
}
\makeatother

\newcolumntype{P}[1]{>{\centering\arraybackslash}p{#1}}

\begin{document}
\title{Faithful Synthesis of Low-dose Contrast-enhanced Brain MRI Scans using Noise-preserving Conditional GANs}
\date{}

\renewcommand{\shorttitle}{Synthesis of Low-dose Contrast-enhanced Brain MRI Scans using CGANs}

%
% \titlerunning{Synthesis of Low-dose Contrast-enhanced Brain MRI Scans using CGANs}
% \author{Thomas Pinetz\inst{1}\orcidID{0000-0002-6100-2136}\thanks{T.P. and A.E. are funded the German Research Foundation under Germany’s Excellence Strategy - EXC-2047/1 - 390685813 and - EXC2151 - 390873048 and R.H. is funded by a research grant (BONFOR; O-194.0002.1).} \and   
%         Erich Kobler\inst{2}\orcidID{0000-0001-5167-4804}\thanks{T. Pinetz and E. Kobler contributed equally to this work.}\and  
%         Robert Haase\inst{2} \orcidID{0000-0003-2311-8004} \and  
%         Katerina Deike-Hofmann\inst{2,3}\orcidID{0000-0002-4319-0428}\and 
%         Alexander Radbruch\inst{2,3}\orcidID{0000-0001-6238-6525}\and  
%         Alexander Effland\inst{1}}  
% index{Pinetz, Thomas} 
% index{Kobler, Erich}
% index{Haase, Robert}
% index{Deike-Hofmann, Katerina}
% index{Radbruch, Alexander}
% index{Effland, Alexander}

\author{
\href{https://orcid.org/0000-0002-6100-2136}{\includegraphics[scale=0.06]{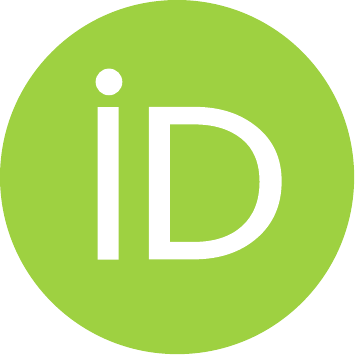}\hspace{1mm}Thomas Pinetz\thanks{Thomas Pinetz and Erich Kobler contributed equally to this work.}}\\
Institute of Applied Mathematics\\
University of Bonn\\
\texttt{pinetz@iam.uni-bonn.de} \\
%% examples of more authors
\And
\href{https://orcid.org/0000-0001-5167-4804}{\includegraphics[scale=0.06]{orcid.pdf}\hspace{1mm}Erich Kobler} \\
Department of Neuroradiology\\
University Medical Center Bonn\\
\texttt{erich.kobler@ukbonn.de} \\
\AND
\href{https://orcid.org/0000-0001-6029-4192}{\includegraphics[scale=0.06]{orcid.pdf}\hspace{1mm}Robert Haase} \\
Department of Neuroradiology\\
University Medical Center Bonn\\
\texttt{robert.haase@ukbonn.de} \\
\And
\href{https://orcid.org/0000-0002-4319-0428}{\includegraphics[scale=0.06]{orcid.pdf}\hspace{1mm}Katerina Deike-Hofmann} \\
Department of Neuroradiology\\
University Medical Center Bonn\\
\texttt{katerina.deike-hofmann@ukbonn.de} \\
\And
\href{https://orcid.org/0000-0001-6238-6525}{\includegraphics[scale=0.06]{orcid.pdf}\hspace{1mm}Alexander Radbruch} \\
Department of Neuroradiology\\
University Medical Center Bonn\\
\texttt{alexander.radbruch@ukbonn.de} \\
\And
Alexander Effland\\
Institute of Applied Mathematics\\
University of Bonn\\
\texttt{effland@iam.uni-bonn.de} \\
}

% \authorrunning{T. Pinetz et al.}
% \institute{Institute of Applied Mathematics, Rheinische Friedrich-Wilhelms-Universität Bonn, Germany \email{\{pinetz, effland\}@iam.uni-bonn.de} \and
% Department of Neuroradiology, University Medical Center Bonn, Germany \and
% German Center for Neurodegenerative Diseases (DZNE), Helmholtz Association of German Research Centers, Germany}
%
\maketitle              % typeset the header of the contribution
\begin{abstract}
Today Gadolinium-based contrast agents (GBCA) are indispensable in Magnetic Resonance Imaging (MRI) for diagnosing various diseases.
However, GBCAs are expensive and may accumulate in patients with potential side effects, thus dose-reduction is recommended.
Still, it is unclear to which extent the GBCA dose can be reduced while preserving the diagnostic value -- especially in pathological regions. 
To address this issue, we collected brain MRI scans at numerous non-standard GBCA dosages and developed a conditional GAN model for synthesizing corresponding images at fractional dose levels.
Along with the adversarial loss, we advocate a novel content loss function based on the Wasserstein distance of locally paired patch statistics for the faithful preservation of noise.
Our numerical experiments show that conditional GANs are suitable for generating images at different GBCA dose levels and can be used to augment datasets for virtual contrast models.
Moreover, our model can be transferred to openly available datasets such as BraTS, where non-standard GBCA dosage images do not exist. Code is available \href{https://github.com/tpinetz/low-dose-gadolinium-mri-synthesis}{here}.
    
\keywords{MRI, GANs, Optimal Transport, Noise Modelling}
\end{abstract}
\section{Introduction}

Magnetic Resonance Imaging (MRI) of the brain is an essential imaging modality to accurately diagnose various neurological diseases ranging from inflammatory lesions to brain tumors and metastases.
For accurate depictions of said pathologies, gadolinium-based contrast agents (GBCA) are injected intravenously to highlight brain-blood barrier dysfunctions.
However, these contrast agents are expensive and may cause nephrogenic systemic fibrosis in patients with severely reduced kidney function~\cite{ScBl18}.
Moreover, \cite{KaIs14} reported that Gadolinium accumulates inside patients with unclear health consequences, especially after repeated application.
The American College of Radiology recommends administering the lowest GBCA dose to obtain the needed clinical information~\cite{ACR22}.
%  a standard dose of 0.1 mmol/kg for clinical usage and recommends further dose reduction~\cite{esur18}.

Driven by this recommendation, several research groups have recently published dose-reduction techniques focusing on maintaining image quality.
Complementary to the development of higher relaxivity contrast agents~\cite{RoPo19}, virtual contrast~\cite{GoPa18,AmBo22} -- replacing a large fraction of the GBCA dose by deep learning -- has been proposed. % TODO: cite our review article once published!
% Recently, several research groups have published work in reducing said dosis by developing higher relaxivity contrast agents~\cite{RoPo19} or by the usage of deep learning~\cite{GoPa18, AmBo22}.
These approaches typically acquire a contrast-enhanced~(CE) scan with a lower GBCA dose along with non-CE scans, e.g., T1w, T2w, FLAIR, or ADC.
These input images are then processed by a deep neural network (DNN) to replicate the corresponding standard-dose scan.
While promising, virtual contrast techniques have not been integrated into clinical practice yet due to false-positive signals or missed small lesions~\cite{LuZh21,AmBo22}.
As with all deep learning-based approaches, the availability of large datasets is essential, which is problematic in the considered case since the additional CE low-dose scan is not acquired in clinical routine exams.
Hence, there are no public datasets to easily benchmark and compare different algorithms or evaluate their performance.
In general, the enhancement behavior of pathological tissues at various GBCA dosages has barely been researched due to a lack of data~\cite{HaPi23b}.

In recent years, generative models have been used to overcome data scarcity in the computer vision and medical imaging community.
Frequently, generative adversarial networks (GANs)~\cite{GoPo14} are applied as state-of-the-art in image generation~\cite{SaSc22} or semantic translation/interpolation~\cite{KaLa19,ArJi20,LiPa22}.
In a nutshell, the GAN framework trains two competing DNNs -- the generator and the discriminator.
The generator learns a non-linear transformation of a predefined noise distribution to fit the distribution of a target dataset, while the discriminator provides feedback by simultaneously approximating a distance or divergence between the generated and the target distribution.
The choice of this distance leads to the well-known different GAN algorithms, e.g., Wasserstein GANs~\cite{ArCh17,GuAh17}, Least Squares GANs~\cite{MaLi17}, or Non-saturating GANs~\cite{GoPo14}.
However, Lucic et al.~\cite{LuKu18} showed that this choice has only a minor impact on the performance.
\begin{figure}[tb]
\centering
\includegraphics[width=.9\linewidth]{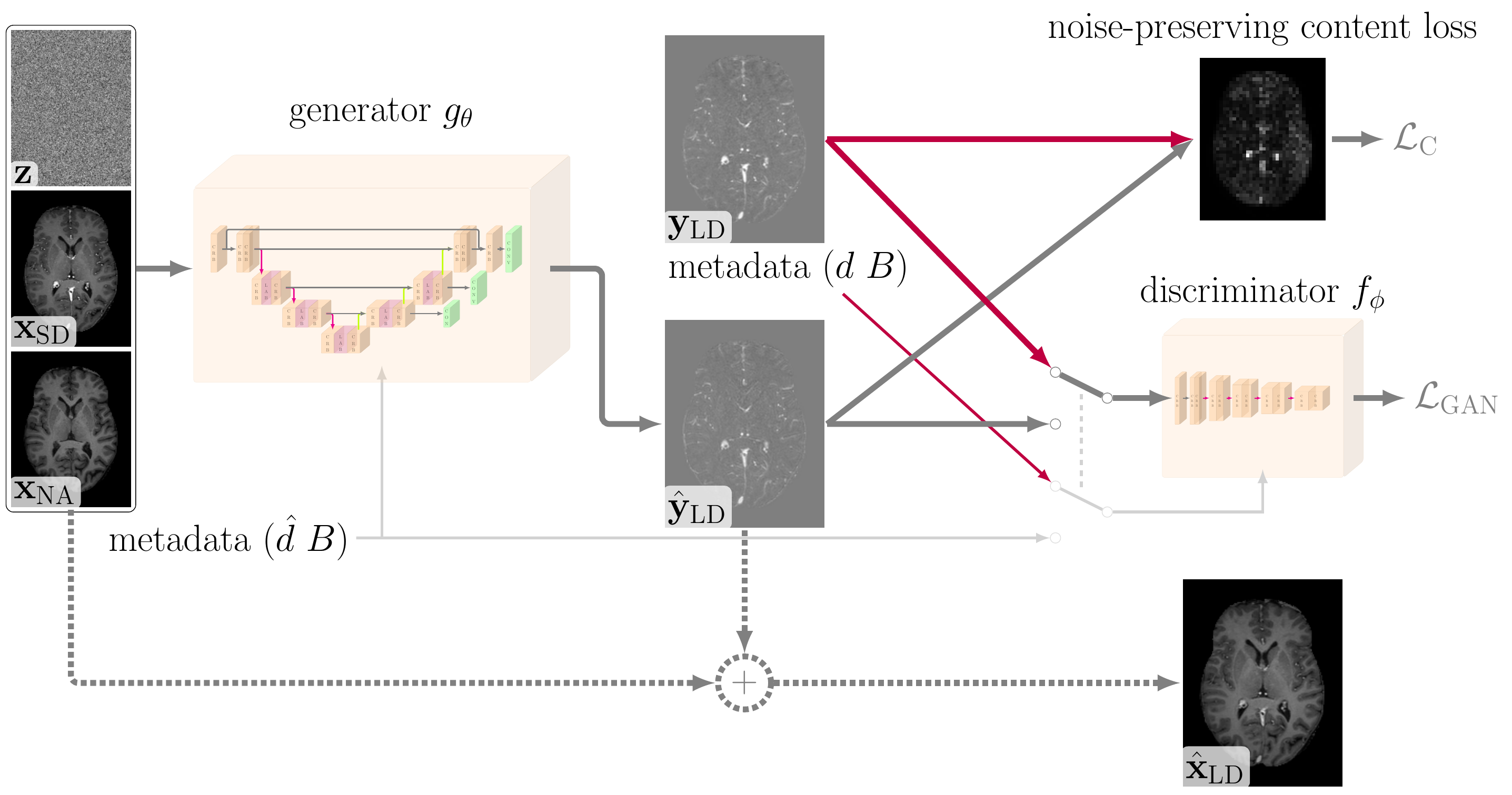}
% \resizebox{\linewidth}{!}{
% \begin{tikzpicture}
% \node at (-12, -3) {\includegraphics[width=1.25\linewidth]{imgs/model.pdf}};
% \foreach \c/\d [count=\x from 0] in {10p/10\%,20p/20\%,33p/33\%} {
% \node at (2.1*\x,0) {\includegraphics[width=0.25\linewidth]{imgs/ProofOfConcept/real_\c.png}};
% \node at (2.1*\x,-2.8) {$\d$};
% \node at (2.1*\x,-5.6) {\includegraphics[width=0.25\linewidth]{imgs/ProofOfConcept/synth_\c.png}};
% }
% \end{tikzpicture}
% }
\caption{Low-dose synthesis using a conditional GAN.
The generator predicts a residual low-dose image~$\hat{\y}_\LD$ from a noise sample~$\z$ conditioned on the native~$\x_\NA$ and standard-dose~$\x_\SD$ images as well as the field strength~$B$ and the artificial dose~$\hat{d}$.
Along with the discriminator, a novel noise-preserving loss -- penalizing the Wasserstein distance of paired patches -- is used for training.
At inference, the generated residual~$\hat{\y}_\LD$ is added to the native image~$\x_\NA$ to yield the corresponding synthetic low-dose~$\hat{\x}_\LD$.}
\label{fig:proofofconcept}
\end{figure}

Learning conditional distributions between images can be accomplished by additionally feeding a condition~(additional scans, dose level, etc.) into both the generator and discriminator.
In particular, for image-to-image translation tasks, these conditional GANs have been successfully applied using paired~\cite{IsZh17,NiTr17,PrMe21} and unpaired training data~\cite{ZhPa17}.
Within these methods, an additional content (cycle) loss typically penalizes pixel-wise deviations (e.g., $\ell_1$) from a corresponding reference to enforce structural similarity, whereas a local adversarial loss (discriminator with local receptive field) controls textural similarity.
In addition, embeddings have been used to inject metadata~\cite{KaLa19,ChUh20}.
% In the original formulation the noise for the adversarial loss was often ignored in favor of a near deterministic output and since then diversity encouraging loss terms have been proposed~\cite{YaHo19,MaLe19}. 

To study the GBCA accumulation behavior, we collected 453~CE scans with non-standard GBCA doses in the set of $\{10\%,20\%,33\%\}$ along with the corresponding standard-dose (0.1 mmol/kg) scan after applying the remaining contrast agent.
Using this dataset, we aim at the semantic interpolation of the GBCA signal at various fractional dose levels.
To this end, we use GANs to learn the contrast enhancement behavior from the dataset collective and thereby enable the synthesis of contrast signals at various dose levels for individual cases.
Further, to minimize the smoothing effect~\cite{LaSo16} of typical content losses (e.g., $\ell_1$ or perceptual~\cite{JoAl16}), we develop a noise-preserving content loss function based on the Wasserstein distance between paired image patches calculated using a Sinkhorn-style algorithm.
This novel loss enables a faithful generation of noise, which is important for the identification of enhancing pathologies and their usability as additional training data.

With this in mind, the contributions of this work are as follows:
\begin{itemize}
\item synthesis of GBCA behavior at various doses using conditional GANs,
\item loss enabling interpolation of dose levels present in training data,
\item noise-preserving content loss function to generate realistic synthetic images.
% \item Visualization of the dynamics of contrast agent enhancement of T1-weighted MRI scans for arbitrary dose concentrations.
\end{itemize}

\section{Methodology}
Given a \emph{native} image~$\x_\NA$ (i.e.~without any contrast agent injection) and a CE \emph{standard-dose} image~$\x_\SD$, our conditional GAN approach synthesizes CE \emph{low-dose} images~$\hat{\x}_\LD$ for selected dose levels~$\hat{d}\in\D\subset[0,1]$ from a uniform noise image~$\z\sim\N(0,\id)$, see Figure~\ref{fig:proofofconcept}.
To focus the generation on the contrast agent signal, our model predicts residual images~$\hat{\y}_\LD$; the corresponding low-dose can be obtained by~$\hat{\x}_\LD=\x_\NA+\hat{\y}_\LD$.

For training and evaluation, we consider samples~$(\x_\NA,\x_\SD,\y_\LD,d,B)$ of a dataset~$\DS$, where $\y_\LD=\x_\LD-\x_\NA$ is the residual image of a real CE low-dose scan~$\x_\LD$ with dose level~$d\in\D$ and $B\in\{1.5,3\}$ is the field-strength in Tesla of the used scanner.
To simplify learning of the contrast accumulation behavior, we adapt the preprocessing pipeline of BraTS~\cite{BaGh21}.
Further details of the dataset and the preprocessing are in the supplementary material.

\subsection{Conditional GANs for Contrast Signal Synthesis}
Our approach is built on the insight that contrast enhancement is an inherently local phenomenon and the necessary information for the synthesis task can be extracted from a local neighborhood within an image.
Therefore, we use as generator~$g_\theta$ a convolutional neural network (CNN) that is based on the U-Net~\cite{RoFi15} along with a local attention mechanism.
The architecture design and the implementation details can be found in the supplementary material.
As illustrated in Figure~\ref{fig:proofofconcept}, the generator uses a 3D noise sample~$\z\sim\N(0,\id)$ along with the native and SD images~$(\x_\NA,\x_\SD)$ as input.
The synthesis is guided by the metadata~$\metad{\hat{d}}{B}$, containing the artificial dose level~$\hat{d}\in\D$ as well as the field strength of the corresponding scanner~$B\in\{1.5,3\}$.
In particular, the metadata is injected into every residual block of the generator using an embedding, motivated by the recent success of diffusion-based models~\cite{HoJa20}.

To learn this generator, a convolutional discriminator~$f_\phi$ is used, which is in turn trained to distinguish the generated residual images~$\hat{\y}_\LD$ with random dose level~$\hat{d}$ from the real residual images~$\y_\LD$ with the associated real dose level~$d$.
To make this a non-trivial task, label smoothing on the metadata is used, i.e, the real dose is augmented by zero-mean additive Gaussian noise with standard deviation~$0.05$.
The discriminator architecture essentially implements the encoding side of the generator, however, no local attention layers are used as suggested by~\cite{LeCh22}.
Like the generator, the discriminator is conditioned on the metadata using an embedding, which is not shared between both networks.

For training of the generator~$\theta$ and discriminator~$\phi$, we consider the loss
\begin{align}
\min_\theta \max_\phi\left\{\L_\mathrm{GAN}(\theta,\phi)+\lambda_\mathrm{GP}\L_\mathrm{GP}(\phi)+\lambda_\mathrm{C}\L_\mathrm{C}(\theta)\right\},
\label{eq:totalLoss}
\end{align}
which consists of a Wasserstein GAN loss~$\L_\mathrm{GAN}$, a gradient penalty loss~$\L_\mathrm{GP}$, and a content loss~$\L_\mathrm{C}$ that are relatively weighted by scalar non-negative factors~$\lambda_\mathrm{GP}$ and $\lambda_\mathrm{C}$.
In detail, the Wasserstein GAN loss reads as
\begin{align*}
    &\L_\mathrm{GAN}(\theta,\phi)\coloneqq\\&\hspace{1em}\E_{(\x_\NA,\x_\SD,\y_\LD,d,B)\sim\U(\DS)}\left\{f_\phi\left(\y_\LD,c\right) - \E_{\z\sim\N(0,\id),\hat{d}\sim\U(\D)}\left\{f_\phi\left(g_\theta\left(\z,\hat{c}\right),\hat{c}\right)\right\}\right\}
\end{align*}
using condition tuples $c=(\x_\NA,\x_\SD,\metad{d}{B})$ and $\hat{c}=(\x_\NA,\x_\SD,\metad{\hat{d}}{B})$ to simplify notation.
$\U(\mathcal{S})$ denotes a uniform distribution over a set~$\mathcal{S}$.
We highlight that the artificial dose levels~$\hat{d}$ for the generated images are uniformly sampled from~$\D=[0.05, 0.5]$, which enables an interpolation around the dose levels present in the dataset~$\DS$.
This is necessary since only a few distinct dose levels~$\{0.1, 0.2, 0.33\}$ have been acquired.
For regularizing the discriminator~$f_\phi$, we include the gradient penalty loss
\[
\L_\mathrm{GP}(\phi)\coloneqq\E_{\subalign{&(\x_\NA,\x_\SD,\y_\LD,d,B)\sim\U(\DS)\\&\z\sim\N(0,\id),\hat{d}\sim\U(\D),\alpha\sim\U(0,1)}}\hspace*{-1em}\left\{(\|\nabla f_\phi(h(\alpha,\y_\LD,\hat{\y}_\LD), h(\alpha,c,\hat{c})) \|_2 - 1)^2\right\}
\]
using $h(\alpha,\y,\hat{\y})=\alpha\hat{\y}+(1-\alpha)\y$.
A penalty term is introduced, if $f_\phi$ is not Lipschitz continuous with factor~$1$ in its arguments as required by Wasserstein GANs~\cite{GuAh17}.
Here,~$\hat{\y}_\LD=g_\theta(\z,\hat{c})$ and the Lipschitz penalty is evaluated at convex combinations of real and synthetic images and condition tuples (essentially dose levels).
Finally, using a distance~$\ell_C$, the content loss
\[
\L_\mathrm{C}(\theta)\coloneqq E_{(\x_\NA,\x_\SD,\y_\LD,d,B)\sim\U(\DS),\z\sim\N(0,\id)}\left\{\ell_C\big(g_\theta\left(\z,c\right), \y_\LD\big)\right\}
\]
guides the generator~$g_\theta$ towards residual images in the dataset.
Thus, it teaches the generator the principles of contrast enhancement.
Typically, the $\ell_1$-norm is used as a distance function, which leads to smooth results since it also penalizes deviations from the noise in~$y_\LD$.

\subsection{Noise-preserving Content Loss}
\begin{figure}[t]
\centering
\includegraphics[width=.95\linewidth]{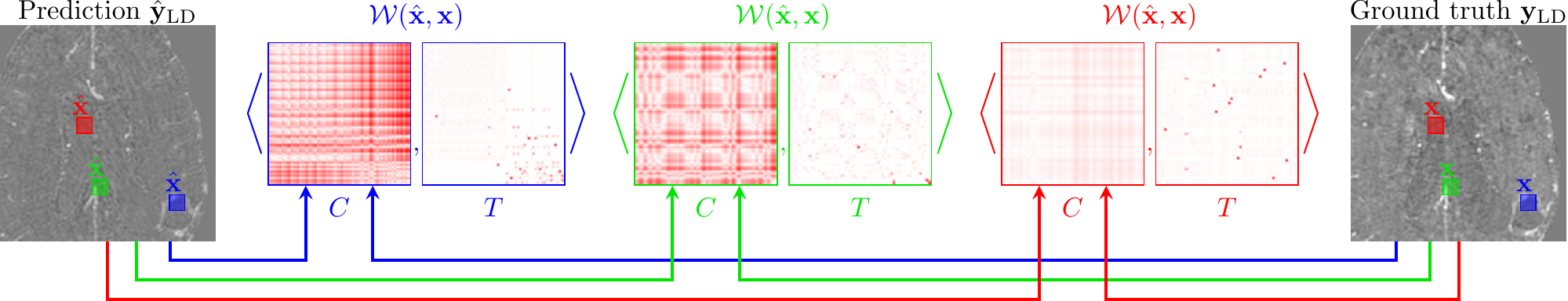}
\caption{
Illustration of our patch-wise noise-preserving content loss.
For each patch pair~($\color{blue}(\hat{\x},\x)$, $\color{green}(\hat{\x},\x)$, $\color{red}(\hat{\x},\x)$) extracted at the same position, the loss accounts for the Wasserstein distance~$\W$ of the associated empirical distributions.
In the center, the corresponding cost matrices~$C$ (pixel-wise absolute difference) along with the optimal transport maps~$T$ are shown, which are obtained by solving~\eqref{eq:WassersteinDistance}.
The final loss is the sum of the element-wise multiplication of all~$C$ and~$T$ for every non-overlapping patch.}
\label{fig:otloss}
\end{figure}

To generate realistic CE images, it is also important to retain the original noise characteristics.
% In this work, we aim at is to realistically estimate the required amount of contrast agent to visualize certain pathologies.
% To do this accurately it is important to retain the original noise structure of the images.
Therefore, we introduce a novel loss that accounts for deviations in local statistics using optimal transport between empirical distributions of paired patches, as illustrated in Figure~\ref{fig:otloss}.
    
Let $\x,\hat{\x}\in\R^{n^3}$ be patches of size~$n\times n\times n$ extracted from the same location of a real and synthetic image, respectively.
The Wasserstein distance of the associated empirical distributions using a transport plan~$T\in\R_+^{n^3\times n^3}$ and cost matrix~$C\in\R_+^{n^3\times n^3}$ given by ($C_{ij}=|\hat{x}_i-x_j|$) is defined as
\begin{align} \label{eq:WassersteinDistance}
    \W(\hat{\x},\x) = \min_{T\in\R_+^{n^3\times n^3}} \langle C, T\rangle_F \qquad \text{s.t.} \quad T\one = \one\tfrac{1}{n^3},\; T^\top\one=\one\tfrac{1}{n^3},
\end{align}
where $\one$ is the vector of ones of size~$n^3$.
In contrast to the element-wise difference penalization of the $\ell_1$-distance, the Wasserstein distance accounts for mismatches between distributions.
To illustrate this, let us, for instance, assume that both patches are Gaussian distributed ($x\sim\N(\mu,\sigma)$, $\hat{x}\sim\N(\hat{\mu},\hat{\sigma})$), which is a coarse simplification of real MRI noise~\cite{AjVe16}.
In this case, the Wasserstein distance reduces to second-order momentum matching, i.e, $\W^2(\hat{\x},\x)=(\mu-\hat{\mu})^2+(\sigma-\hat{\sigma})^2$.
Thus, the Wasserstein distance generalizes this distributional loss to any distribution within paired patches.

To efficiently solve problem~\eqref{eq:WassersteinDistance}, we use the inexact proximal point algorithm~\cite{XiXi20}.
This algorithm is parallelized and applied to all non-overlapping patch pairs, to obtain our noise-preserving content loss
\[
\ell_\mathrm{NP}(\hat{\y},\y) = \E_{\off\sim\U(\O)}\Big\{\sum_{\pat\in\P} \W(P_{\pat+\off}\hat{\y},P_{\pat+\off}\y)\Big\},
\]
where $P_{\pat}$ extracts a local $n\times n\times n$ patch at location~$\pat\in\P=\{0,n,2n,\ldots\}^3$ using periodic boundary conditions.
Note that we compute the expectation over offsets~$\off\in\O=\{0,1,\ldots,\lceil\tfrac{n}{2}\rceil\}^3$ to avoid patching artifacts.
In the numerical implementation, only a single offset is sampled for time and memory constraints.

\section{Numerical Results}
\label{sec:results}
In this section, we evaluate the proposed conditional GAN approach with a particular focus on different content loss distance functions.
All synthesis models were trained on 250 samples acquired on 1.5T and 3T Philips Achieva scanners and evaluated on 193 test cases, all collected at site~\site{1}.
Further details of the dataset, model and training can be found in the supplementary.
In our experiments, we observed that the choice of the content loss distance function~$\ell_C(\hat{\y},\y)$ strongly influences the performance.
Thus, we consider the different cases:\\[1ex]
\hspace*{1cm}$\ell_1:\quad\|\hat{\y}-\y\|_1$ \hfill
VGG:$\quad\|h(\hat{\y})-h(\y)\|_1$ \hfill
NP:$\quad\ell_\mathrm{NP}(\hat{\y},\y)$\hspace*{1cm}\\[1ex]
% \begin{center}
% \begin{tabular}{l | c | c | c}
%                         & $\ell_1$            & VGG                       &  NP \\ \hline
% $\ell_C(\hat{\y},\y) =$ & $\|\hat{\y}-\y\|_1$ & $\|h(\hat{\y})-h(\y)\|_1$ & $\ell_\mathrm{NP}(\hat{\y},\y)$
% \end{tabular}
% \end{center}
Following Johnsen et al.~\cite{JoAl16}, $h(\x)$ is the VGG-16 model~\cite{SiZi15} up to \texttt{relu3\_3}.

\begin{figure}[tbh]
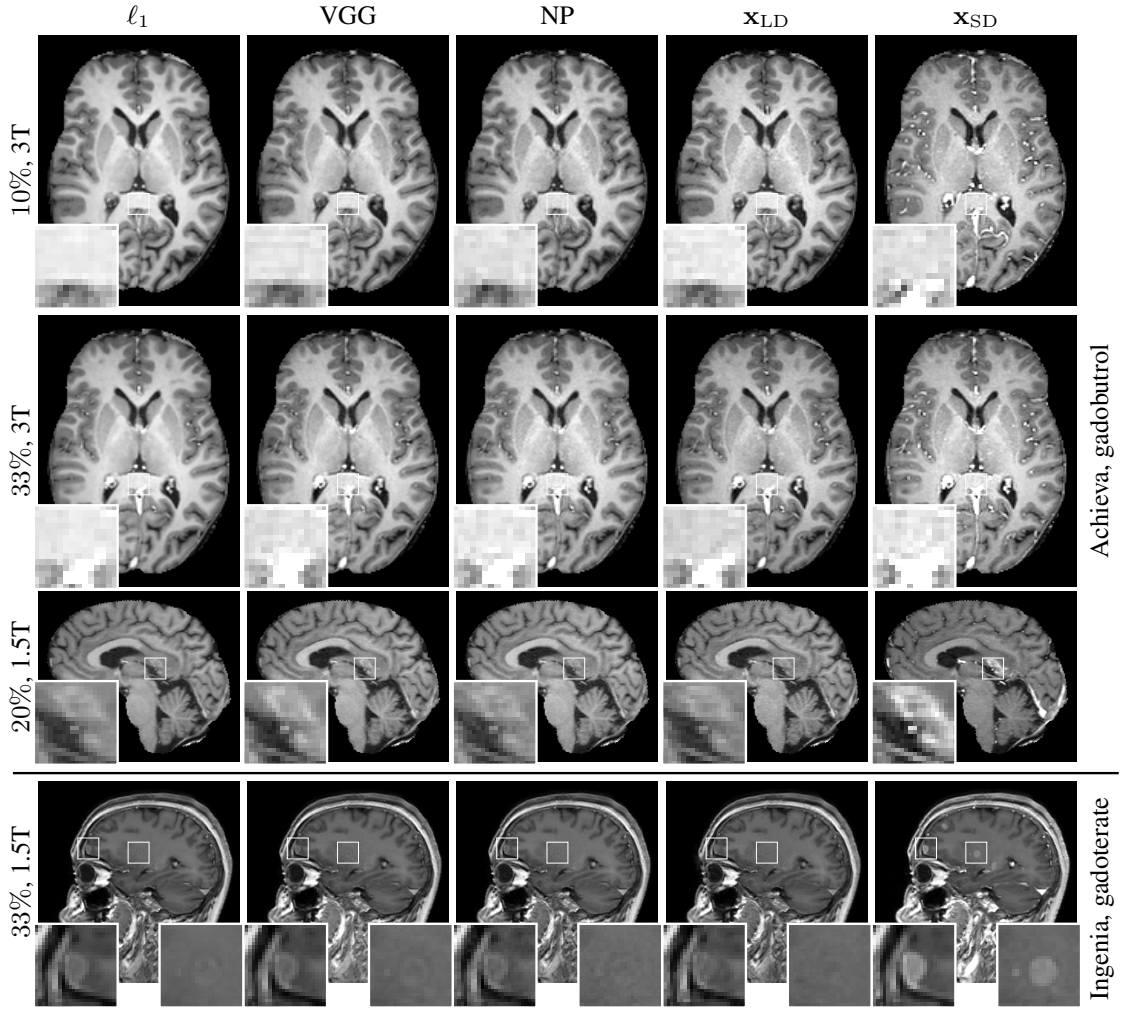

  \centering
  \resizebox{.9\linewidth}{!}{
  \begin{tikzpicture}[spy using outlines={white,magnification=4, size=1cm},
                      every node/.append style={inner sep=0,font=\small}]
  \foreach \d [count=\x from 0] in {$\ell_1$,VGG,NP,$\mathbf{x}_\LD$,$\mathbf{x}_\SD$} {
    \node[anchor=base] at (2.5*\x,1.75) {\d};
  }
  \foreach \c [count=\x from 0] in {10p_l1,10p_vgg,10p_wloss,10p_low,full_dose} {
    \node at (2.5*\x, 0) {\includegraphics[width=2.4cm]{imgs/diff_losses/VC_000000097/contrast/\c_VC5.png}};
  }
  \spy on (2.5*0+.0,0-.4) in node at (2.5*0-.75,-1.15);
  \spy on (2.5*1+.0,0-.4) in node at (2.5*1-.75,-1.15);
  \spy on (2.5*2+.0,0-.4) in node at (2.5*2-.75,-1.15);
  \spy on (2.5*3+.0,0-.4) in node at (2.5*3-.75,-1.15);
  \spy on (2.5*4+.0,0-.4) in node at (2.5*4-.75,-1.15);
  \node[rotate=90] at (-1.4,-0) {$10\%$, 3T};
  
  \foreach \c [count=\x from 0] in {10p_l1,10p_vgg,10p_wloss,10p_low,full_dose} {
    \node at (2.5*\x, -3.35) {\includegraphics[width=2.4cm]{imgs/diff_losses/VC_000000612/contrast/\c_VC5.png}};
  }
  \spy on (2.5*0+.0,-3.35-.4) in node at (2.5*0-.75,-3.35-1.15);
  \spy on (2.5*1+.0,-3.35-.4) in node at (2.5*1-.75,-3.35-1.15);
  \spy on (2.5*2+.0,-3.35-.4) in node at (2.5*2-.75,-3.35-1.15);
  \spy on (2.5*3+.0,-3.35-.4) in node at (2.5*3-.75,-3.35-1.15);
  \spy on (2.5*4+.0,-3.35-.4) in node at (2.5*4-.75,-3.35-1.15);
  \node[rotate=90] at (-1.4,-3.35) {$33\%$, 3T};
  
  \foreach \c [count=\x from 0] in {10p_l1,10p_vgg,10p_wloss,10p_low,full_dose} {
    \node at (2.5*\x,-6.05) {\includegraphics[width=2.4cm]{imgs/diff_losses/VC_000000232/contrast/\c_VC5.png}};
  }
  \spy on (2.5*0+.2,-6.05+.1) in node at (2.5*0-.75,-6.05-.55);
  \spy on (2.5*1+.2,-6.05+.1) in node at (2.5*1-.75,-6.05-.55);
  \spy on (2.5*2+.2,-6.05+.1) in node at (2.5*2-.75,-6.05-.55);
  \spy on (2.5*3+.2,-6.05+.1) in node at (2.5*3-.75,-6.05-.55);
  \spy on (2.5*4+.2,-6.05+.1) in node at (2.5*4-.75,-6.05-.55);
  \node[rotate=90] at (-1.4,-6.05) {$20\%$, 1.5T};
  
  % \foreach \c [count=\x from 0] in {l1,vgg,wloss,low} {
  %   \node at (2.5*\x, -8.6) {\includegraphics[width=2.4cm]{imgs/diff_losses/VC_000000613/10p_\c_VC5.png}};
  % }
  % \spy on (2.5*0+.3,-8.6+.3) in node at (2.5*0-.75,-8.6-1.);
  % \spy on (2.5*1+.3,-8.6+.3) in node at (2.5*1-.75,-8.6-1.);
  % \spy on (2.5*2+.3,-8.6+.3) in node at (2.5*2-.75,-8.6-1.);
  % \spy on (2.5*3+.3,-8.6+.3) in node at (2.5*3-.75,-8.6-1.);
  % \node[rotate=90] at (-1.4,-8.6) {$33\%$, 3T};
  
  \foreach \c [count=\x from 0] in {l1_pred,vgg_pred,pred,low_dose, full_dose} {
    \node at (2.5*\x, -8.5) {\includegraphics[width=2.4cm]{imgs/DiffScanner/contrast/\c_arad.png}};
  }
  \spy on (2.5*0-.6,-8.5+.4) in node at (2.5*0-.75,-8.5-1.);
  \spy on (2.5*1-.6,-8.5+.4) in node at (2.5*1-.75,-8.5-1.);
  \spy on (2.5*2-.6,-8.5+.4) in node at (2.5*2-.75,-8.5-1.);
  \spy on (2.5*3-.6,-8.5+.4) in node at (2.5*3-.75,-8.5-1.);
  \spy on (2.5*4-.6,-8.5+.4) in node at (2.5*4-.75,-8.5-1.);

  \spy on (2.5*0-.0,-8.5+.35) in node at (2.5*0+.75,-8.5-1.);
  \spy on (2.5*1-.0,-8.5+.35) in node at (2.5*1+.75,-8.5-1.);
  \spy on (2.5*2-.0,-8.5+.35) in node at (2.5*2+.75,-8.5-1.);
  \spy on (2.5*3-.0,-8.5+.35) in node at (2.5*3+.75,-8.5-1.);
  \spy on (2.5*4-.0,-8.5+.35) in node at (2.5*4+.75,-8.5-1.);
  \node[rotate=90] at (-1.4,-8.5) {$33\%$, 1.5T};

  \node[rotate=90] at (11.5,-3.35) {\footnotesize Achieva, gadobutrol};
  \draw[thick] (-1.5,-7.2) -- (11.7,-7.2);
  \node[rotate=90] at (11.5,-8.7) {\footnotesize Ingenia, gadoterate};

  \end{tikzpicture}}
\caption{Qualitative comparison of synthesized images using different loss functions to the corresponding reference~$\x_\LD$.
While the $\ell_1$ loss yields smooth low-dose images, the noise pattern is preserved to some extent using the VGG loss; our loss helps to further retain the noise characteristics.}
\label{fig:diffloss}
\end{figure}
A qualitative comparison of the different distance functions~$\ell_C$ is visualized in Figure~\ref{fig:diffloss}.
The first column depicts synthesized images using the $\ell_1$-norm as the distance function.
These images depict a plausible contrast signal, however, suffer from unrealistic smooth homogeneous regions.
An improvement thereof is shown by the perceptual content loss (VGG).
The NP-loss leads to a further improvement not only in the contrast signal behavior but also in the realism of the noise texture, cf. zoom regions in the lower corners.

To highlight the generalization capabilities, we depict in the bottom row of Figure~\ref{fig:diffloss} a sample from site~\site{2}, which was acquired using a Philips Ingenia scanner.
Moreover, the GBCA gadoterate was used, while our training data only consists of scans using the GBCA gadobutrol.
Nevertheless, all models present realistically synthesized LD images.
Comparing the zooms of the LD images, we observe that our NP-loss leads to a better synthesis of noise and thereby to more realistic LD images.
In the $\ell_1$ and VGG columns, the noise is not faithfully synthesized, thus it is visually easy to spot the enhancing pathological regions.

For completeness, a quantitative ablation of the considered distance functions on the test images of site~\site{1} is shown in Table~\ref{tab:quantitative}.
Although neither maximizing PSNR nor SSIM~\cite{WaBo04} is our objective, we observe on-par performances of the perceptual~(VGG) and our proposed content loss~(NP) with the standard~$\ell_1$ distance function.
Using the SD image, we define CE pixels as those pixels at which the intensity increases by at least~$10\%$ compared to the native scan.
An example of these CE regions is illustrated in the supplementary.
Thus, the mean absolute error for CE pixels~(MAE$_\mathrm{CE}$) quantifies the enhancement behavior.
Further, we estimate the standard deviation of the non-CE pixels and report the MAE to the ground truth standard deviation~(MAE$_\sigma$).
As shown in Table~\ref{tab:quantitative}, our loss outperforms the other content losses to a large extent on both metrics, proving its effectiveness for faithful contrast enhancement and noise generation.
Further statistical analyses are presented in the supplementary.

Next, we evaluate the effect of synthesized LD images on the performance of a virtual contrast model (VCM).
In particular, we consider the state-of-the-art 2.5D U-Net model~\cite{PaJo21,LuZh21,HaPi23}, which predicts an SD image given a corresponding native and LD image, see supplementary for further details.
The columns on the right of Table~\ref{tab:quantitative} list the average PSNR and SSIM score on the \emph{real} 33\% LD subset of our test data from site~\site{1}. 
The bottom row depicts the performance if just \emph{real} 33\% LD images are used for training the VCM as an upper bound.
In contrast, the other entries on the right list the performance if \emph{only synthesized} LD images are used for training.
Both metrics show that the samples synthesized using our NP-loss model are superior to both~$\ell_1$ and VGG.

To determine the effectiveness of the LD synthesis models at different settings, we acquired 160 data samples from 1.5T and 3T Philips Ingenia scanners at site~\site{2}.
This site used the GBCA gadoterate, which has a lower relaxivity compared to gadobutrol used at site~\site{1}~\cite{JaDu10}.
For 80 samples real LD images were acquired, which are used for testing.
Using the VCM solely trained on the real 33\% LD data of site~\site{1} yields an average PSNR and MAE$_\mathrm{CE}$ on the test samples of site~\site{2} of 40.04 and 0.092, respectively.
Extending the training data for the VCM by synthesized LD images from our model with NP-loss, we get a significantly improvemed (p < 0.001) PSNR score of 40.37 and MAE$_\mathrm{CE}$ of 0.075.

\begin{figure}[hbt]
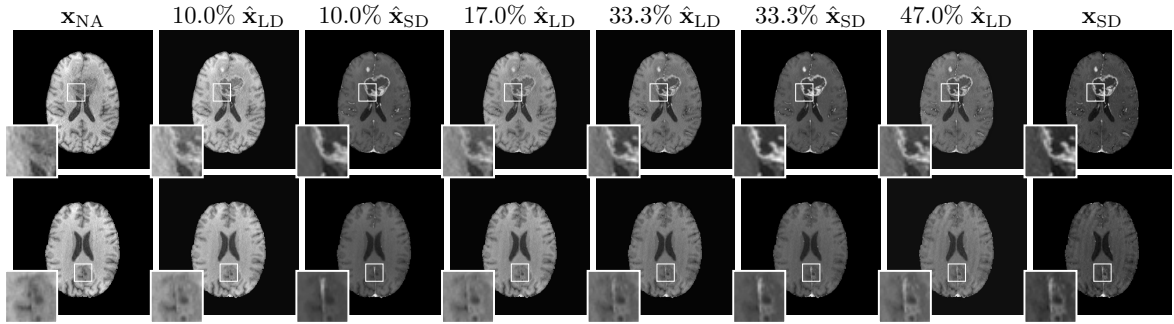

\centering
\resizebox{.95\linewidth}{!}{
\begin{tikzpicture}[spy using outlines={white,magnification=3,size=.75cm},
                    every node/.append style={inner sep=0}]
% \node at (-7, -2) {
%   \begin{tabular}{P{1.25cm}|P{1.25cm}P{1.25cm}}
%         & PSNR  & SSIM\\\hline
%   10.0\%& 37.73 & 0.977 \\
%   20.0\%& 38.53 & 0.983 \\
%   33.3\%& 40.98 & 0.988
%   \end{tabular}
%   };
\foreach \d [count=\x from 0] in {\mathbf{x}_\NA,10.0\%\;\hat{\mathbf{x}}_\LD,10.0\%\;\hat{\mathbf{x}}_\SD,17.0\%\;\hat{\mathbf{x}}_\LD,33.3\%\;\hat{\mathbf{x}}_\LD,33.3\%\;\hat{\mathbf{x}}_\SD,47.0\% \;\hat{\mathbf{x}}_\LD,\mathbf{x}_\SD} {
\node[anchor=base] at (2.1*\x,1.1) {$\d$};
}
\foreach \c [count=\x from 0] in {T1_zero,T1_low_10p,T1_full_10p,T1_low_17p,T1_low_30p,T1_full_30p,T1_low_47p,T1_full} {
\node at (2.1*\x,0) {\includegraphics[height=2cm]{imgs/brats/results_reduction/working_case/\c.png}};
}
\spy on (2.1*0-.1,0+.1) in node at (2.1*0-.75,-.75);
\spy on (2.1*1-.1,0+.1) in node at (2.1*1-.75,-.75);
\spy on (2.1*2-.1,0+.1) in node at (2.1*2-.75,-.75);
\spy on (2.1*3-.1,0+.1) in node at (2.1*3-.75,-.75);
\spy on (2.1*4-.1,0+.1) in node at (2.1*4-.75,-.75);
\spy on (2.1*5-.1,0+.1) in node at (2.1*5-.75,-.75);
\spy on (2.1*6-.1,0+.1) in node at (2.1*6-.75,-.75);
\spy on (2.1*7-.1,0+.1) in node at (2.1*7-.75,-.75);
\foreach \c [count=\x from 0] in {T1_zero,T1_low_10p,T1_full_10p,T1_low_17p,T1_low_30p,T1_full_30p,T1_low_47p,T1_full} {
\node at (2.1*\x,-2.1) {\includegraphics[width=2cm]{imgs/brats/results_reduction/fail_case/\c.png}};
}
\spy on (2.1*0+0.,-2.1-.4) in node at (2.1*0-.75,-2.1-.75);
\spy on (2.1*1+0.,-2.1-.4) in node at (2.1*1-.75,-2.1-.75);
\spy on (2.1*2+0.,-2.1-.4) in node at (2.1*2-.75,-2.1-.75);
\spy on (2.1*3+0.,-2.1-.4) in node at (2.1*3-.75,-2.1-.75);
\spy on (2.1*4+0.,-2.1-.4) in node at (2.1*4-.75,-2.1-.75);
\spy on (2.1*5+0.,-2.1-.4) in node at (2.1*5-.75,-2.1-.75);
\spy on (2.1*6+0.,-2.1-.4) in node at (2.1*6-.75,-2.1-.75);
\spy on (2.1*7+0.,-2.1-.4) in node at (2.1*7-.75,-2.1-.75);
\end{tikzpicture}
}
\caption{Comparison of synthesized LD images~$\hat{\x}_\LD$ and corresponding predicted SD images~$\hat{\x}_\SD$ for different dose levels on BraTS~\cite{BaGh21} along with the native (left) and real SD image (right).
We also included non-fractional dosage levels (17\% and 47\%) to showcase the wide applicability of our algorithm.
Top: the tumor is well contrasted in all~$\hat{\x}_\SD$ even for 10\%.
Bottom: the subtle enhancement of the tumor cannot be recovered from the 10\% LD image.}
\label{fig:bratsresults}
\end{figure}

Finally, Figure~\ref{fig:bratsresults} visualizes synthesized LD images on the BraTS dataset~\cite{BaGh21} along with the associated VCM outputs.
Comparing the predicted SD images~$\hat{\x}_\SD$ using 10\% and 33\% synthesized LD images~$\hat{\x}_\LD$, we observe that the weakly enhancing tumor at the bottom zoom is not preserved in the case of 10\%, enabling evaluation of dose reduction methods on known pathological regions.

\begin{table}[t]
\centering
\caption{Quantitative comparison of the low-dose synthesis methods.
The central columns present metrics evaluated on the synthesized low-dose images, whereas the right columns evaluate the effect of purely synthesized data for training the standard-dose prediction model~\cite{PaJo21}.
Note, that the PSNR/SSIM of the standard dose prediction model was always evaluated on real LD images.
The definitions of the mean absolute error on the contrast enhancement (MAE$_\mathrm{CE}$) and on the noise standard deviation~(MAE$_\sigma$) are in Section~\ref{sec:results}.
A $^\ast$ denotes if a Wilcoxon signed rank test between VGG and NP{\tiny (our)} row is significant.
}
\label{tab:quantitative}%
\begin{tabular}{p{1.25cm}|P{1.25cm}P{1.25cm}P{1.25cm}P{1.25cm}|P{1.75cm}P{1.75cm}}
                 & \multicolumn{4}{c|}{low-dose synthesis}                        & \multicolumn{2}{c}{standard-dose prediction}\\\cline{2-7}
                 & PSNR                  & SSIM            & MAE$_\mathrm{CE}$    & MAE$_\sigma$         & PSNR                 & SSIM  \\ \hline
$\ell_1$         & \textbf{38.34}        & \textbf{0.978}  & 0.022                & 0.012                & 33.83                & 0.922 \\
VGG              & 37.84                 & \textit{0.976}  & \textit{0.019}       & \textit{0.009}       & \textit{36.33}       & \textit{0.958} \\
NP{\tiny (our)}  & \textit{38.05}$^\ast$ & \textit{0.976}  & \textbf{0.011}$^\ast$& \textbf{0.004}$^\ast$& \textbf{37.15}$^\ast$& \textbf{0.960}$^\ast$ \\\hline
$\mathbf{x}_\LD${\tiny (real)}&                       &                 &                      &         & 39.07                & 0.974 \\ 
\end{tabular}
\end{table}

\section{Conclusions}

In this work, we used conditional GANs to synthesize contrast-enhanced images using non-standard GBCA doses.
To this end, we introduced a novel noise-preserving content loss motivated by optimal transport theory.
Numerous numerical experiments showed that our content loss improves the faithful synthesis of low-dose images.
Further, the performance of virtual contrast models increases if training data is extended by synthesized images from our GAN model trained by the noise-preserving content loss.

\section*{Acknowledgements}
Thomas Pinetz and Alexander Effland are funded the German Research Foundation under Germany’s Excellence Strategy - EXC-2047/1 - 390685813 and - EXC2151 - 390873048.
Robert Haase is funded by a research grant (BONFOR; O-194.0002.1).


\bibliographystyle{unsrtnat}

\begin{thebibliography}{35}
\providecommand{\natexlab}[1]{#1}
\providecommand{\url}[1]{\texttt{#1}}
\expandafter\ifx\csname urlstyle\endcsname\relax
  \providecommand{\doi}[1]{doi: #1}\else
  \providecommand{\doi}{doi: \begingroup \urlstyle{rm}\Url}\fi

\bibitem[Schieda et~al.(2018)Schieda, Blaichman, Costa, Glikstein, Hurrell,
  James, Jabehdar~Maralani, Shabana, Tang, Tsampalieros, B.~van~der Pol, and
  Hiremath]{ScBl18}
Nicola Schieda, Jason~I Blaichman, Andreu~F Costa, Rafael Glikstein, Casey
  Hurrell, Matthew James, Pejman Jabehdar~Maralani, Wael Shabana, An~Tang, Anne
  Tsampalieros, Christian B.~van~der Pol, and Swapnil Hiremath.
\newblock Gadolinium-based contrast agents in kidney disease: a comprehensive
  review and clinical practice guideline issued by the canadian association of
  radiologists.
\newblock \emph{Canadian Journal of Kidney Health and Disease}, 5, 2018.

\bibitem[Kanda et~al.(2014)Kanda, Ishii, Kawaguchi, Kitajima, and
  Takenaka]{KaIs14}
Tomonori Kanda, Kazunari Ishii, Hiroki Kawaguchi, Kazuhiro Kitajima, and
  Daisuke Takenaka.
\newblock High signal intensity in the dentate nucleus and globus pallidus on
  unenhanced t1-weighted mr images: relationship with increasing cumulative
  dose of a gadolinium-based contrast material.
\newblock \emph{Radiology}, 270\penalty0 (3):\penalty0 834--841, 2014.

\bibitem[ACR(2022)]{ACR22}
\emph{ACR Manual on Contrast Media}.
\newblock American College of Radiology, 2022.
\newblock ISBN 978-1-55903-012-0.

\bibitem[Robic et~al.(2019)Robic, Port, Rousseaux, Louguet, Fretellier, Catoen,
  Factor, Le~Greneur, Medina, Bourrinet, Raynal, Id{\'e}e, and Corot]{RoPo19}
Caroline Robic, Marc Port, Olivier Rousseaux, St{\'e}phanie Louguet, Nathalie
  Fretellier, Sarah Catoen, C{\'e}cile Factor, Soizic Le~Greneur, Christelle
  Medina, Philippe Bourrinet, Isabelle Raynal, Jean-Marc Id{\'e}e, and Claire
  Corot.
\newblock Physicochemical and pharmacokinetic profiles of gadopiclenol: a new
  macrocyclic gadolinium chelate with high t1 relaxivity.
\newblock \emph{Investigative Radiology}, 54\penalty0 (8):\penalty0 475, 2019.

\bibitem[Gong et~al.(2018)Gong, Pauly, Wintermark, and Zaharchuk]{GoPa18}
Enhao Gong, John~M Pauly, Max Wintermark, and Greg Zaharchuk.
\newblock Deep learning enables reduced gadolinium dose for contrast-enhanced
  brain {MRI}.
\newblock \emph{Journal of Magnetic Resonance Imaging}, 48\penalty0
  (2):\penalty0 330--340, 2018.

\bibitem[Ammari et~al.(2022)Ammari, B{\^o}ne, Balleyguier, Moulton, Chouzenoux,
  Volk, Menu, Bidault, Nicolas, Robert, et~al.]{AmBo22}
Samy Ammari, Alexandre B{\^o}ne, Corinne Balleyguier, Eric Moulton, {\'E}milie
  Chouzenoux, Andreas Volk, Yves Menu, Fran{\c{c}}ois Bidault, Fran{\c{c}}ois
  Nicolas, Philippe Robert, et~al.
\newblock Can deep learning replace gadolinium in neuro-oncology?: a reader
  study.
\newblock \emph{Investigative Radiology}, 57\penalty0 (2):\penalty0 99--107,
  2022.

\bibitem[Luo et~al.(2021)Luo, Zhang, Gong, Tamir, Venkata, Xu, Duan, Zhou,
  Zhou, Zaharchuk, Xue, and Liu]{LuZh21}
Huanyu Luo, Tao Zhang, Nan-Jie Gong, Jonthan Tamir, Srivathsa~Pasumarthi
  Venkata, Cheng Xu, Yunyun Duan, Tao Zhou, Fuqing Zhou, Greg Zaharchuk, Jing
  Xue, and Yaou Liu.
\newblock Deep learning--based methods may minimize gbca dosage in brain mri.
\newblock \emph{European Radiology}, 31\penalty0 (9):\penalty0 6419--6428,
  2021.

\bibitem[Haase et~al.(2023{\natexlab{a}})Haase, Pinetz, Kobler, Paech, Effland,
  Radbruch, and Deike-Hofmann]{HaPi23b}
Robert Haase, Thomas Pinetz, Erich Kobler, Daniel Paech, Alexander Effland,
  Alexander Radbruch, and Katerina Deike-Hofmann.
\newblock Artificial contrast: Deep learning for reducing gadolinium-based
  contrast agents in neuroradiology.
\newblock \emph{Investigative Radiology}, 2023{\natexlab{a}}.

\bibitem[Goodfellow et~al.(2014)Goodfellow, Pouget-Abadie, Mirza, Xu,
  Warde-Farley, Ozair, Courville, and Bengio]{GoPo14}
Ian Goodfellow, Jean Pouget-Abadie, Mehdi Mirza, Bing Xu, David Warde-Farley,
  Sherjil Ozair, Aaron Courville, and Yoshua Bengio.
\newblock Generative adversarial nets.
\newblock In Z.~Ghahramani, M.~Welling, C.~Cortes, N.~Lawrence, and K.Q.
  Weinberger, editors, \emph{NeurIPS}, volume~27. Curran Associates, Inc.,
  2014.

\bibitem[Sauer et~al.(2022)Sauer, Schwarz, and Geiger]{SaSc22}
Axel Sauer, Katja Schwarz, and Andreas Geiger.
\newblock Stylegan-xl: Scaling stylegan to large diverse datasets.
\newblock In \emph{ACM SIGGRAPH}, pages 1--10, 2022.

\bibitem[Karras et~al.(2019)Karras, Laine, and Aila]{KaLa19}
Tero Karras, Samuli Laine, and Timo Aila.
\newblock A style-based generator architecture for generative adversarial
  networks.
\newblock In \emph{CVPR}, pages 4401--4410, 2019.

\bibitem[Armanious et~al.(2020)Armanious, Jiang, Fischer, K{\"u}stner, Hepp,
  Nikolaou, Gatidis, and Yang]{ArJi20}
Karim Armanious, Chenming Jiang, Marc Fischer, Thomas K{\"u}stner, Tobias Hepp,
  Konstantin Nikolaou, Sergios Gatidis, and Bin Yang.
\newblock Medgan: Medical image translation using gans.
\newblock \emph{Computerized Medical Imaging and Graphics}, 79:\penalty0
  101684, 2020.

\bibitem[Liu et~al.(2022)Liu, Pasumarthi, Duffy, Gong, Zaharchuk, and
  Datta]{LiPa22}
Jiang Liu, Srivathsa Pasumarthi, Ben Duffy, Enhao Gong, Greg Zaharchuk, and
  Keshav Datta.
\newblock One model to synthesize them all: Multi-contrast multi-scale
  transformer for missing data imputation.
\newblock \emph{arXiv preprint arXiv:2204.13738}, 2022.

\bibitem[Arjovsky et~al.(2017)Arjovsky, Chintala, and Bottou]{ArCh17}
Martin Arjovsky, Soumith Chintala, and L{\'e}on Bottou.
\newblock Wasserstein generative adversarial networks.
\newblock In \emph{ICML}, pages 214--223, 2017.

\bibitem[Gulrajani et~al.(2017)Gulrajani, Ahmed, Arjovsky, Dumoulin, and
  Courville]{GuAh17}
Ishaan Gulrajani, Faruk Ahmed, Martin Arjovsky, Vincent Dumoulin, and Aaron~C
  Courville.
\newblock Improved training of wasserstein gans.
\newblock \emph{NeurIPS}, 30, 2017.

\bibitem[Mao et~al.(2017)Mao, Li, Xie, Lau, Wang, and Paul~Smolley]{MaLi17}
Xudong Mao, Qing Li, Haoran Xie, Raymond~YK Lau, Zhen Wang, and Stephen
  Paul~Smolley.
\newblock Least squares generative adversarial networks.
\newblock In \emph{ICCV}, pages 2794--2802, 2017.

\bibitem[Lucic et~al.(2018)Lucic, Kurach, Michalski, Gelly, and
  Bousquet]{LuKu18}
Mario Lucic, Karol Kurach, Marcin Michalski, Sylvain Gelly, and Olivier
  Bousquet.
\newblock Are gans created equal? a large-scale study.
\newblock In \emph{NeurIPS}, 2018.

\bibitem[Isola et~al.(2017)Isola, Zhu, Zhou, and Efros]{IsZh17}
Phillip Isola, Jun-Yan Zhu, Tinghui Zhou, and Alexei~A. Efros.
\newblock Image-to-image translation with conditional adversarial networks.
\newblock In \emph{CVPR}, pages 1125--1134, 2017.

\bibitem[Nie et~al.(2017)Nie, Trullo, Lian, Petitjean, Ruan, Wang, and
  Shen]{NiTr17}
Dong Nie, Roger Trullo, Jun Lian, Caroline Petitjean, Su~Ruan, Qian Wang, and
  Dinggang Shen.
\newblock Medical image synthesis with context-aware generative adversarial
  networks.
\newblock In \emph{MICCAI}, pages 417--425, 2017.

\bibitem[Preetha et~al.(2021)Preetha, Meredig, Brugnara, Mahmutoglu, Foltyn,
  Isensee, Tobias, Pfl{\"u}ger, Schell, Neuberger, et~al.]{PrMe21}
Chandrakanth~Jayachandran Preetha, Hagen Meredig, Gianluca Brugnara, Mustafa~A
  Mahmutoglu, Martha Foltyn, Fabian Isensee, Kessler Tobias, Irada Pfl{\"u}ger,
  Marianne Schell, Ulf Neuberger, et~al.
\newblock Deep-learning-based synthesis of post-contrast t1-weighted mri for
  tumour response assessment in neuro-oncology: a multicentre, retrospective
  cohort study.
\newblock \emph{The Lancet Digital Health}, 3\penalty0 (12):\penalty0
  e784--e794, 2021.

\bibitem[Zhu et~al.(2017)Zhu, Park, Isola, and Efros]{ZhPa17}
Jun-Yan Zhu, Taesung Park, Phillip Isola, and Alexei~A Efros.
\newblock Unpaired image-to-image translation using cycle-consistent
  adversarial networks.
\newblock In \emph{ICCV}, pages 2223--2232, 2017.

\bibitem[Choi et~al.(2020)Choi, Uh, Yoo, and Ha]{ChUh20}
Yunjey Choi, Youngjung Uh, Jaejun Yoo, and Jung-Woo Ha.
\newblock Stargan v2: Diverse image synthesis for multiple domains.
\newblock In \emph{CVPR}, pages 8188--8197, 2020.

\bibitem[Larsen et~al.(2016)Larsen, S{\o}nderby, Larochelle, and
  Winther]{LaSo16}
Anders Boesen~Lindbo Larsen, S{\o}ren~Kaae S{\o}nderby, Hugo Larochelle, and
  Ole Winther.
\newblock Autoencoding beyond pixels using a learned similarity metric.
\newblock In \emph{ICML}, pages 1558--1566, 2016.

\bibitem[Johnson et~al.(2016)Johnson, Alahi, and Fei-Fei]{JoAl16}
Justin Johnson, Alexandre Alahi, and Li~Fei-Fei.
\newblock Perceptual losses for real-time style transfer and super-resolution.
\newblock In \emph{ECCV}, pages 694--711. Springer, 2016.

\bibitem[Baid et~al.(2021)Baid, Ghodasara, Mohan, Bilello, Calabrese, Colak,
  Farahani, Kalpathy-Cramer, Kitamura, Pati, et~al.]{BaGh21}
Ujjwal Baid, Satyam Ghodasara, Suyash Mohan, Michel Bilello, Evan Calabrese,
  Errol Colak, Keyvan Farahani, Jayashree Kalpathy-Cramer, Felipe~C Kitamura,
  Sarthak Pati, et~al.
\newblock The rsna-asnr-miccai brats 2021 benchmark on brain tumor segmentation
  and radiogenomic classification.
\newblock \emph{arXiv preprint arXiv:2107.02314}, 2021.

\bibitem[Ronneberger et~al.(2015)Ronneberger, Fischer, and Brox]{RoFi15}
Olaf Ronneberger, Philipp Fischer, and Thomas Brox.
\newblock U-net: Convolutional networks for biomedical image segmentation.
\newblock In \emph{MICCAI}, pages 234--241. Springer, 2015.

\bibitem[Ho et~al.(2020)Ho, Jain, and Abbeel]{HoJa20}
Jonathan Ho, Ajay Jain, and Pieter Abbeel.
\newblock Denoising diffusion probabilistic models.
\newblock \emph{NeurIPS}, 33:\penalty0 6840--6851, 2020.

\bibitem[Lee et~al.(2022)Lee, Chang, Jiang, Zhang, Tu, and Liu]{LeCh22}
Kwonjoon Lee, Huiwen Chang, Lu~Jiang, Han Zhang, Zhuowen Tu, and Ce~Liu.
\newblock Vi{TGAN}: Training {GAN}s with vision transformers.
\newblock In \emph{ICLR}, 2022.

\bibitem[Aja-Fern{\'a}ndez and Vegas-S{\'a}nchez-Ferrero(2016)]{AjVe16}
Santiago Aja-Fern{\'a}ndez and Gonzalo Vegas-S{\'a}nchez-Ferrero.
\newblock \emph{Statistical analysis of noise in MRI}.
\newblock Springer, 2016.

\bibitem[Xie et~al.(2020)Xie, Wang, Wang, and Zha]{XiXi20}
Yujia Xie, Xiangfeng Wang, Ruijia Wang, and Hongyuan Zha.
\newblock A fast proximal point method for computing exact wasserstein
  distance.
\newblock In \emph{Uncertainty in Artificial Intelligence}, pages 433--453,
  2020.

\bibitem[Simonyan and Zisserman(2015)]{SiZi15}
Karen Simonyan and Andrew Zisserman.
\newblock Very deep convolutional networks for large-scale image recognition.
\newblock In \emph{ICLR}, pages 1--14, 2015.

\bibitem[Wang et~al.(2004)Wang, Bovik, Sheikh, and Simoncelli]{WaBo04}
Zhou Wang, Alan~C Bovik, Hamid~R Sheikh, and Eero~P Simoncelli.
\newblock Image quality assessment: from error visibility to structural
  similarity.
\newblock \emph{Transactions on Image Processing}, 13\penalty0 (4):\penalty0
  600--612, 2004.

\bibitem[Pasumarthi et~al.(2021)Pasumarthi, Tamir, Christensen, Zaharchuk,
  Zhang, and Gong]{PaJo21}
Srivathsa Pasumarthi, Jonathan~I Tamir, Soren Christensen, Greg Zaharchuk, Tao
  Zhang, and Enhao Gong.
\newblock A generic deep learning model for reduced gadolinium dose in
  contrast-enhanced brain mri.
\newblock \emph{Magnetic Resonance in Medicine}, 86\penalty0 (3):\penalty0
  1687--1700, 2021.

\bibitem[Haase et~al.(2023{\natexlab{b}})Haase, Pinetz, Bendella, Kobler,
  Paech, Block, Effland, Radbruch, and Deike-Hofmann]{HaPi23}
Robert Haase, Thomas Pinetz, Zeynep Bendella, Erich Kobler, Daniel Paech,
  Wolfgang Block, Alexander Effland, Alexander Radbruch, and Katerina
  Deike-Hofmann.
\newblock Reduction of gadolinium-based contrast agents in mri using
  convolutional neural networks and different input protocols: Limited
  interchangeability of synthesized sequences with original full-dose images
  despite excellent quantitative performance.
\newblock \emph{Investigative Radiology}, 2023{\natexlab{b}}.

\bibitem[Jacques et~al.(2010)Jacques, Dumas, Sun, Troughton, Greenfield, and
  Caravan]{JaDu10}
Vincent Jacques, St{\'e}phane Dumas, Wei-Chuan Sun, Jeffrey~S Troughton,
  Matthew~T Greenfield, and Peter Caravan.
\newblock High relaxivity mri contrast agents part 2: Optimization of inner-and
  second-sphere relaxivity.
\newblock \emph{Investigative Radiology}, 45\penalty0 (10):\penalty0 613, 2010.

\end{thebibliography}

\pagebreak

\section*{Supplementary material}

\subsection*{Dataset Details}

\begin{table*}
\begin{tabular}{c| c | c| c | c | c | c | c | c}
Device & Male/Female & Weight & Field strength & Time to Echo & Repetition time & Flip angle & Resolution (mm) \\ \hline
Achieva & 175/131 & $81.5 \pm 18.3$ &  3 & $3.1 \pm 0.1$ & $6.8 \pm 0.1$ & 8 & $0.5\times0.5\times0.5$ \\
Achieva & 78/69 & $78.5 \pm 17.1$  & 1.5 & $3.3 \pm 0.1$ & $7.3 \pm 0.1$ & 8 & $0.5\times0.5\times0.5$ \\
Ingenia & 22/7 & $ 90.3 \pm 20.2  $ & 3. & $3.8\pm 0.1$ & $7,7 \pm 0.1$ & 8 & $0.5\times0.5\times1.0$ \\
Ingenia & 71/51 & $84.5 \pm 19.3 $  & 1.5 & $3.6\pm 0.0$ & $15.4 \pm 0.04$ & 30 & $0.75\times0.75\times1.0$ \\
\end{tabular}
\caption{Details on the acquired data for this study}
\label{tab:dataset_details}
\end{table*}

\begin{figure}[htb]
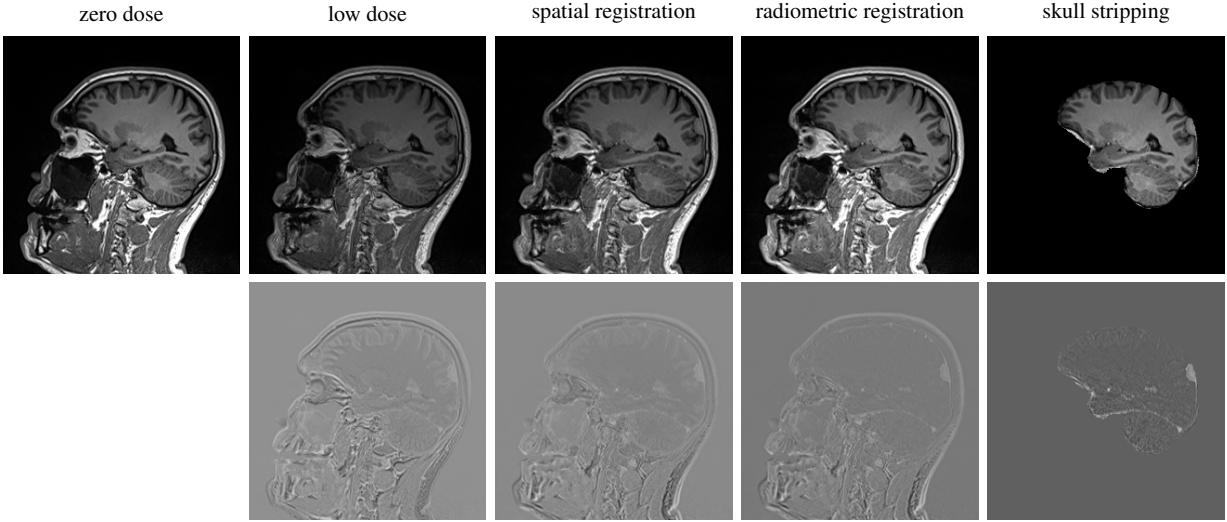

  \centering
  \resizebox{\linewidth}{!}{
  \begin{tikzpicture}
    \foreach \c [count=\x from 0] in {zero dose,low dose,spatial registration,radiometric registration,skull stripping} {
      \node at (3.9*\x,2.25) {\c};
    }
    \foreach \c [count=\x from 0] in {zero_dose,low_dose_nonreg,low_dose_nonradreg,low_dose, low_dose_mask} {
      \node at (3.9*\x,0) {\includegraphics[height=3.75cm]{imgs/preprocessing/\c.png}};
    }
    \foreach \c [count=\x from 1] in {low_diff_nonreg,low_diff_nonradreg,low_diff, low_diff_mask} {
      \node at (3.9*\x,-3.9) {\includegraphics[height=3.75cm]{imgs/preprocessing/\c.png}};
    }
  \end{tikzpicture}}
  \caption{Visualization of the preprocessing steps. The top row shows the zero dose and the registration steps applied to a low-dose image.
  We adapt the preprocessing pipeline of BraTS~\cite{BaGh21}, i.e, resampling, normalization, spatial registration, skull stripping, and radiometric registration.
  The bottom row illustrates the differences between the processed low-dose image and the zero dose.}
  \label{fig:Preprocessing}
\end{figure}

Our dataset consists of 453 studies, which were collected following approval by the Ethics Committee for Clinical Trials on Humans and Epidemiological Research with Personal Data of the Faculty of Medicine of the Rheinische Friedrich-Wilhelms-Universität Bonn (reference no. 450/20).
Details are given in Table \ref{tab:dataset_details}
Our training dataset uses a stratified split and consists of $50$~patients with the following field strength and dose level combinations: 3T/0.1, 3T/0.2, 3T/0.33, 1.5T/0.2, and 1.5T/0.33.
10 patients are used to validate the algorithm and tune hyperparameters and the remaining 193 patients are used for testing.
Additional 151 patients have been acquired at two different scanner to show the generalizability of our method and is detailed in Table \ref{tab:dataset_details}.
The preprocessing pipeline thereof is shown in Fig.\ref{fig:Preprocessing}.

\begin{figure}[htb]
  \centering
  \includegraphics[width=.95\linewidth]{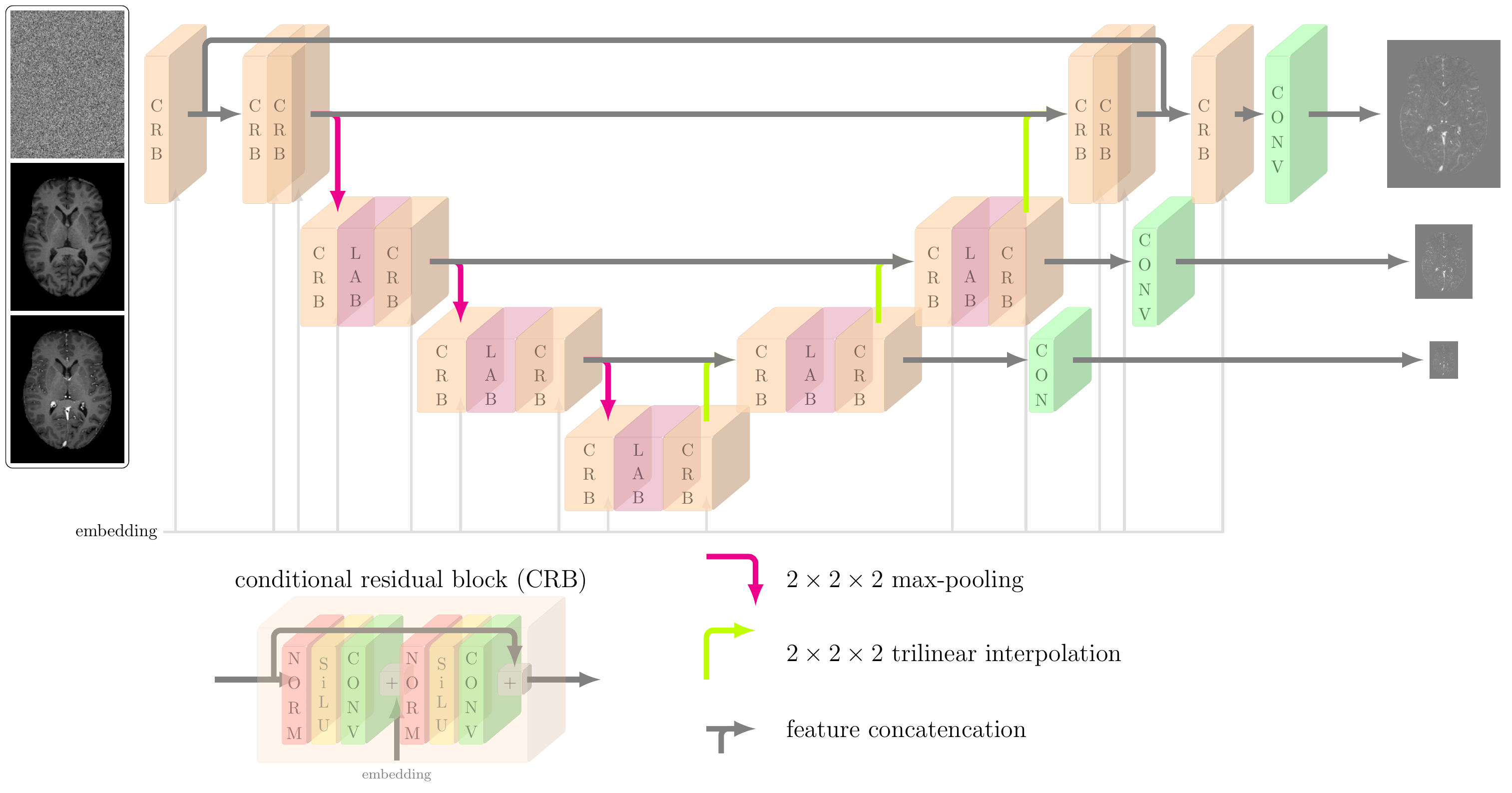}
  \caption{Architecture of the generator~$g_\theta$ following a U-Net approach.
  At all four scales, conditional residual blocks (CRBs) are used to locally extract the contrast behavior from the native~$\x_\NA$ and standard-dose~$\x_\SD$ images.}
  \label{fig:Generator}
  \end{figure}

\subsection*{Implementation Details}

The architecture of the generator network~$g_\theta$ is depicted in Figure~\ref{fig:Generator} and is implemented using pytorch (1.13).
At all four scales, conditional residual blocks (CRBs) are used to locally extract the contrast behavior from the native~$\x_\NA$ and standard-dose~$\x_\SD$ images.
Every CRB receives a linear projection of an embedding that is generated by feeding the metadata into a two-layer neural network using~$128$ features.
Every linear projection layer is learnable.
At the finest scale, $12$~feature channels are extracted, which are doubled after each max-pooling operation.
The coarser scales additionally use local attention blocks (LABs) to aggregate local information within a $3\times3\times3$ search window using self-attention.
All convolution layers utilize $3\times3\times3$~kernels.

As described in the paper, the discriminator~$f_\phi$ implements the encoding side of the generator but used 5 scales and no LABs.
The hyperparameters are identical to the generator.

To increase the receptive field of our noise-preserving loss~$\ell_\mathrm{NP}$, we output multiple heads in the generator, see Figure~\ref{fig:Generator}.
Then, $\ell_\mathrm{NP}$ is computed in every scale using patches of size~$n=4$.
For solving the Wasserstein problem, we perform 100 iterations of the inexact proximal point algorithm~\cite{XiXi20} to estimate the optimal transport map for every patch pair using the regularization parameter~$\epsilon=\tfrac {\max(C)} {10}$.

For the optimization of the saddle point problem, the Adam~\cite{KiBa15} optimizer was used for both the generator~$g_\theta$ and the discriminator~$f_\phi$ with learning rates of $10^{-4}$ and momentum variables $(0.5, 0.9)$.
$5$~discriminator updates are performed for each generator update.
The training is performed for $100.000$ iterations, with a minibatch size of~$12$ and a patch size of $64^3$ on a A100 GPU with a runtime of 2 days and 13 hours.
The weighting of the loss terms is independent of the used loss and we always set $\lambda_\mathrm{C} = 1$ and $\lambda_\mathrm{GAN} = 1$. 
The gradient penalty term $\lambda_\mathrm{GP}$ is set to $100$ for stable training.
For the virtual contrast model, the hyperparameters are taken from~\cite{PaJo21}, with the only difference being the initial image size, which was set to 256 to reflect the changes in resolution from 0.5 mm to 1.0 mm.
An ablation study on sharing the embedding between the generator and the discriminator is shown in Tab.\ref{tab:ablation}.
Having a separate embedding for the discriminator and generator gives a slight performance improvement across all metrics.
The p-values for the mean absolute error (MAE) in the contrast-enhanced regions~(MAE$_\mathrm{CE}$) and MAE of estimated standard deviations in non-contrast-enhanced regions~(MAE$_\sigma$) indicate the difference is statistically significant.

\begin{table}[t]
\centering
\caption{Ablation study between having shared or separate embeddings using PSNR, SSIM, and mean absolute error on contrast-enhanced regions~(MAE$_\mathrm{CE}$), and MAE of estimated standard deviations in non-contrast-enhanced regions~(MAE$_\sigma$). p-values are computed using Wilcoxon signed rank test between GAN+VGG and GAN+Our.}%
\label{tab:ablation}%
\begin{tabular}{l|cccc}
Metric & Shared embeddings  & Separate embeddings \\ \hline
PSNR &  37.85 & \textbf{38.05} \\
SSIM & {0.975} & \textbf{0.976} \\
MAE$_\mathrm{CE}$ & {0.015} & \textbf{0.011} (p $< 0.001$) \\
MAE$_\sigma$ & {0.006} & \textbf{0.004} (p $< 0.001$)\\
\end{tabular}
\end{table}

\subsection*{Statistical evaluation of the contrast enhancement}
\label{appendix:plausibility}

\begin{figure}[htb]
  \centering
  \resizebox{\linewidth}{!}{
  \begin{tikzpicture}
  \foreach \c [count=\x from 0] in {nativ,full,sub,km_mask} {
  \node at (3.9*\x,0) {\includegraphics[height=5cm]{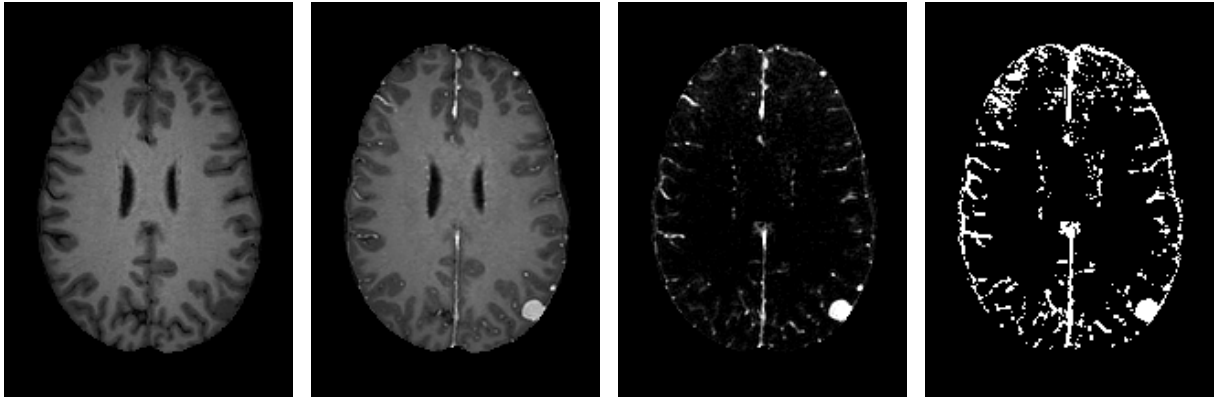}};
  }
  \end{tikzpicture}
  }
\caption{From left to right: native~$\mathbf{x}_\NA$, standard dose~$\mathbf{x}_\SD$, subtraction~$\mathbf{x}_\SD-\mathbf{x}_\NA$ and the CE mask generated by thresholding the subtraction image by~$0.1$, which is equivalent to a $10\%$ contrast increase compared to the $0.95$-quantile of the native image~$\mathbf{x}_\NA$.}
\label{fig:km_mask}
\end{figure}

\begin{figure}[htb]
\centering
\resizebox{\linewidth}{!}{
\begin{tikzpicture}
% \centering
% \foreach \c [count=\x from 0] in {nativ,full} {
% \node at (3.*\x + 11.5,0) {\includegraphics[height=4cm]{imgs/km_mask/\c.png}};
% }
% \foreach \c [count=\x from 0] in {sub,km_mask} {
% \node at (3.*\x + 11.5,-5) {\includegraphics[height=4cm]{imgs/km_mask/\c.png}};
% }
\draw[thick] (9.5,2.5) -- (9.5,-7.5);
\node at (0, 0) {\includegraphics[width=6cm]{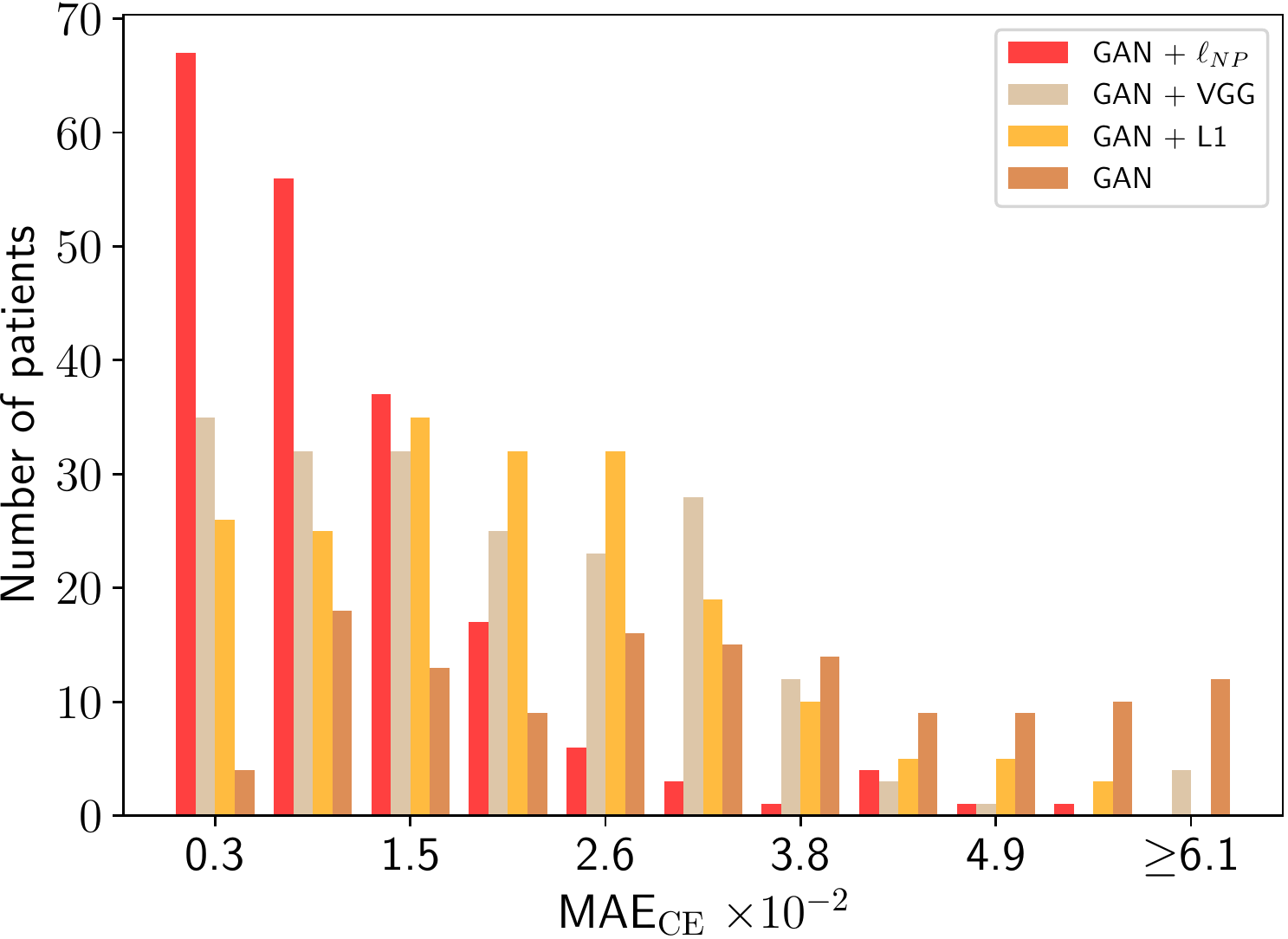}};
\node at (6.25, 0) {\includegraphics[width=6cm]{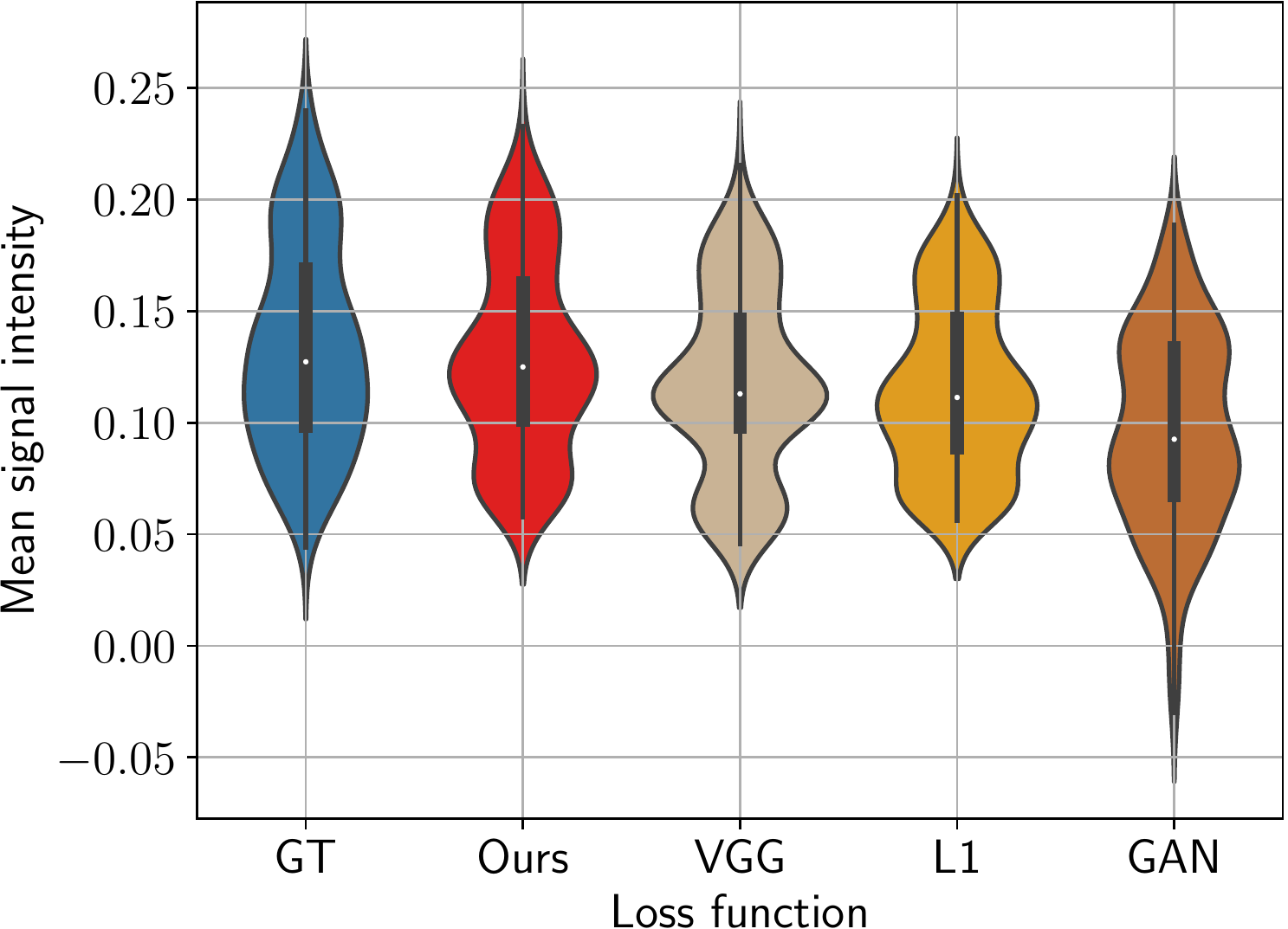}};
\draw[thick] (-3,-2.8) -- (9.5,-2.8);
\node at (0, -5.5) {\includegraphics[width=6cm]{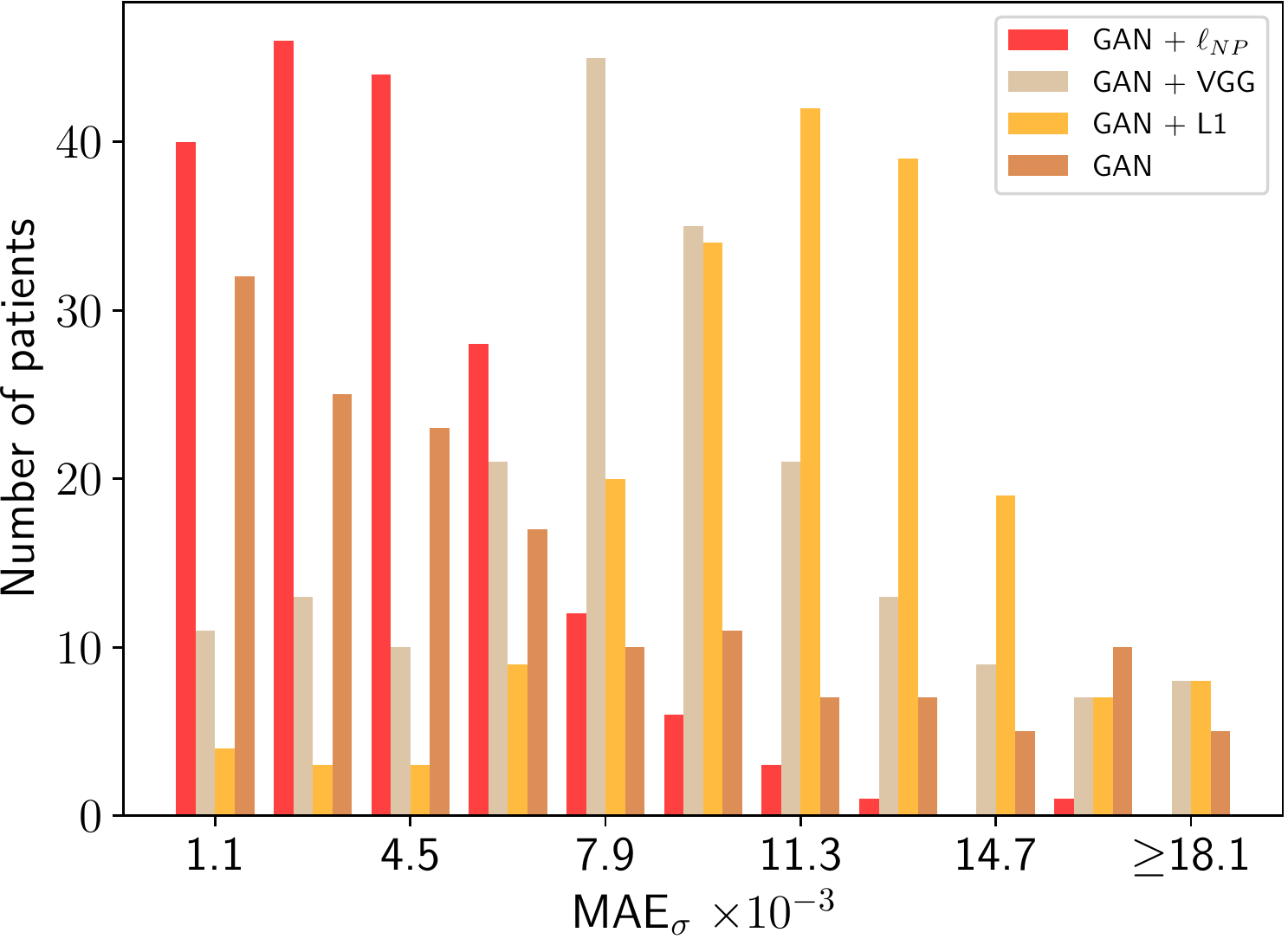}};
\node at (6.25, -5.5) {\includegraphics[width=6cm]{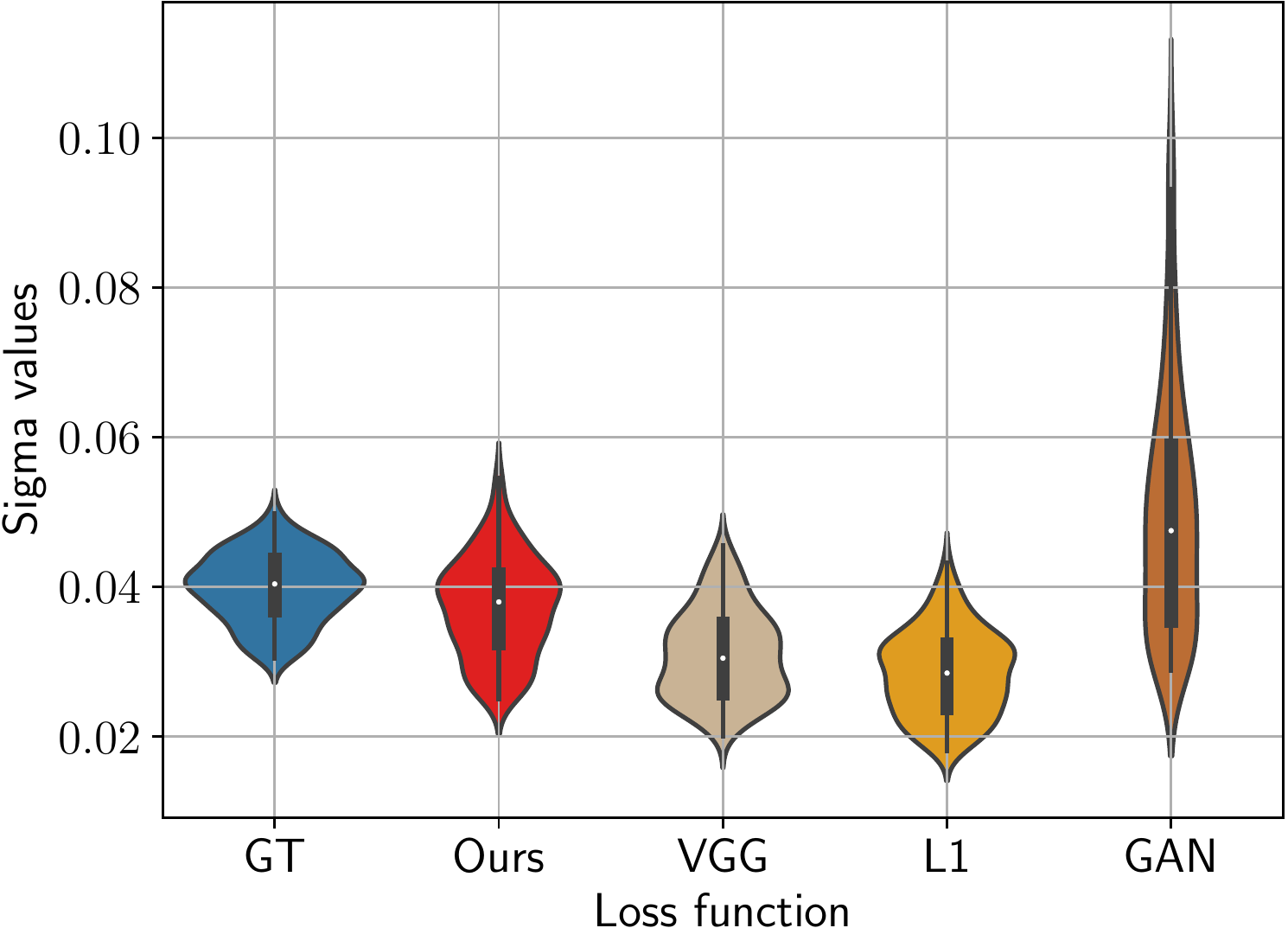}};
\end{tikzpicture}}
\caption{
Statistical evaluation of contrast-enhancing regions (top) and non-enhancing regions (bottom).
The histograms on the left compare the considered loss functions in both regions. While the top histogram shows the mean absolute error in the contrast signal, the bottom histogram visualizes the mean absolute error of the estimated standard deviations.
The proposed noise preserving loss~$\ell_{NP}$ has smaller densities at low errors in both metrics.
The violin plots on the right illustrate how well the statistics within the CE region and non-CE regions overlap with the ground truth (GT).
Comparing the different colored areas, we see that our distance function leads to the best overlap in both regions.}
\label{fig:plausibility}
\end{figure}

In this section, the contrast enhancement of the generated images is statistically validated.
For this reason, the CE pixels are extracted using the corresponding residual image of the standard-dose CE image toward the native image.
For simplicity, we assume that every pixel that enhances at least by 10 $\%$ of our 0.95-quantile is contrast affected, while we consider the remainder as noise.
The binary mask of CE pixels for a typical sample is illustrated in Figure~\ref{fig:km_mask}.
The contrast enhancement is then calculated using the mean over the CE pixels.
Histograms of the mean absolute intensity errors in CE pixels (top) and the mean absolute standard deviation error in non-CE pixels (bottom) for the different loss formulations are found on the left in Figure~\ref{fig:plausibility}.
The violin plots on the right illustrate the signal distribution in CE (top) and the standard deviations in non-CE (bottom) regions.
The plots show that the contrast enhancement of our synthetic images better fits the ground truth low-dose images and the distribution of the contrast enhancements of our loss formulation better matches the ground truth than the other content loss formulations.
The same observation holds for the standard deviation estimation in non-CE regions.

\end{document}